\documentclass[runningheads]{llncs}
\usepackage{graphicx}
\usepackage{CJKutf8}

\begin{document}
\title{ICDAR 2021 Competition on Scientific Literature Parsing}
%
%
\author{Antonio Jimeno Yepes\inst{1,2} \and
Peter Zhong\inst{1,3} \and
Douglas Burdick\inst{4}}
\authorrunning{A. Jimeno Yepes et al.}
%
\institute{IBM Research, Australia \and
University of Melbourne, Australia \\
\email{antonio.jimeno@unimelb.edu.au}\\
 \and
 Oracle, Melbourne, Australia \\
\email{peter.zhong@oracle.com}
 \and
 IBM Research Almaden, USA \\
\email{drburdic@us.ibm.com}
}
\maketitle              
%
\setcounter{footnote}{0}

\begin{abstract}

Scientific literature contain important information related to cutting-edge innovations in diverse domains. Advances in natural language processing have been driving the fast development in automated information extraction from scientific literature. However, scientific literature is often available in unstructured PDF format. While PDF is great for preserving basic visual elements, such as characters, lines, shapes, etc., on a canvas for presentation to humans, automatic processing of the PDF format by machines presents many challenges.  With over 2.5 trillion PDF documents in existence, these issues are prevalent in many other important application domains as well.

A critical challenge for automated information extraction from scientific literature is that documents often contain content that is not in natural language, such as figures and tables. Nevertheless, such content usually illustrates key results, messages, or summarizations of the research. To obtain a comprehensive understanding of scientific literature, the automated system must be able to recognize the layout of the documents and parse the non-natural-language content into a machine readable format. 

Our ICDAR 2021 Scientific Literature Parsing Competition (ICDAR2021-SLP) aims to drive the advances specifically in document understanding.
ICDAR2021-SLP leverages the PubLayNet and PubTabNet datasets, which provide hundreds of thousands of training and evaluation examples.
In Task A, Document Layout Recognition, submissions with the highest performance combine object detection and specialised solutions for the different categories.  
In Task B, Table Recognition, top submissions rely on methods to identify table components and post-processing methods to generate the table structure and content.
Results from both tasks show an impressive performance and opens the possibility for high performance practical applications.


\keywords{Document Layout Understanding  \and Table Recognition \and ICDAR competition.}
\end{abstract}

\section{Introduction}

Documents in Portable Document Format (PDF) are ubiquitous with over 2.5 trillion documents~\cite{staar2018corpus} available from several industries, including insurance documents to medical files to peer-review scientific articles. PDF represents one of the main sources of knowledge both online and offline. While PDF is great for preserving the basic elements (characters, lines, shapes, images, etc.) on a canvas for different operating systems or devices for humans to consume, it’s not a format that machines can understand.

Most of the current methods for document understanding rely on deep learning, which requires a large number of training examples. We have generated large data sets automatically using PubMed Central\footnote{\url{https://www.ncbi.nlm.nih.gov/pmc}} that have been used in this competition.
PubMed Central is a large collection of full text articles in the biomedical domain provided by the US NIH/National Library of Medicine.

As of today, PubMed Central has almost 7 million full text articles from 2,476 journals, which offers the possibility to study the problem of document understanding over a large set of different article styles.
Our data set has been generated using a subset of PubMed Central that is distributed under a Creative Commons license available for commercial use.

The competition is split in two tasks that address the understanding of document layouts by asking participants to identify several categories of information in document pages (Task A) and the understanding of tables by asking participants to produce an HTML version of table images (Task B).
The IBM Research AI Leaderboard system was used to collect and evaluate the submissions of the participants. This system is based on EvalAI\footnote{\url{https://eval.ai/}}.
In task A, participants had access to all the data except for the ground truth of the final evaluation test set, the test set was released when PubLayNet was made available.
In task B, we released the final evaluation test set three days before submitting the final result by the participants.

We had a large number of participant submissions with 281 for the Evaluation Phase of Task A from 78 different teams.
Results from both tasks show an impressive performance by current state-of-the-art algorithms, improving significantly over previously reported results, which opens the possibility for high performance practical applications.

\section{Task A - Document Layout Recognition}

This task aims to advance the research in recognizing the layout of unstructured documents. Participants of this competition need to develop a model that can identify the common layout elements in document images, including text, titles, tables, figures, and lists, with confidence score for each detection.
The competition site is available from~\footnote{\url{Task A website: https://aieval.draco.res.ibm.com/challenge/41/overview}}.

\subsection{Related work}

There has been several competitions for document layout understanding, with many organised as ICDAR competitions.
Examples of these competitions include~\cite{antonacopoulos2009realistic}, which cover as well complex layouts~\cite{clausner2015enp,clausner2017icdar2017}, which are limited in size.
There are as well data sets for document layout understanding outside competitions, for example the US NIH National Library of Medicine Medical Article Records Groundtruth (MARG) that was obtained from scanned article pages.

Overall, the previous data sets available for document layout understanding are of limited size, typically several hundred pages. The main reason for the limited size is that ground-truth data is annotated manually, which is a slow, costly, tedious process.
In our Task A competition, we provide a significantly larger data set, several orders of magnitude larger, that has been generated automatically in which the validation and test sets have been manually verified.

\subsection{Data}

This task used the PubLayNet dataset\footnote{https://github.com/ibm-aur-nlp/PubLayNet}~\cite{zhong2019publaynet}. The annotations in PubLayNet are automatically generated by matching the PDF format and the XML format of the articles in the PubMed Central Open Access Subset as described in ~\cite{zhong2019publaynet}.

The competition had two phases. The Format Verification Phase spanned the entire competition, for participants to verify their results file met our submission requirements with the provided mini development set. The Evaluation Phase also spanned the whole competition. In this phase, participants could submit results on the test samples for evaluation. Final ranking and winning teams were decided by the performance in the Evaluation Phase.
Table~\ref{tab:task-a-data} shows the statistics of the data sets used in the different phases of the Task A competition.

\begin{table}[htpb!]
\begin{center}
\begin{tabular}{|l|r|l|}
\hline
Split           & Size   &Phase \\
\hline
Training        & 335,703&N/A \\
Development     & 11,245 &N/A \\
Mini development& 20     &Format Verification Phase \\
Test            & 11,405 &Evaluation \\
\hline
\end{tabular}
\end{center}
\label{tab:task-a-data}
\caption{Task A data set statistics}
\end{table}

The results submitted by the participants have been objectively and quantitatively evaluated using the mean average precision (MAP) @ intersection over union (IoU) [0.50:0.95] metric on bounding boxes, which is used in the COCO object detection competition\footnote{http://cocodataset.org/\#detection-eval}.
We calculated the average precision for a sequence of IoU thresholds ranging from 0.50 to 0.95 with a step size of 0.05.
Then, the mean of the average precision on all element categories was computed as the final score.

\subsection{Results}

In the Evaluation Phase, we had more than 200 submissions from over 80 teams.
Table~\ref{tab:task-a-results} shows the top 9 results for the Evaluation Phase of the competition.
Overall results and individual results are significantly higher compared to previously reported results~\cite{zhong2019publaynet}.
The three top systems manage to have an overall performance above 0.97.

The top performing systems, as described in the next section, relied on object detection approaches, which is similar to previous work on this data set.
In addition, the predictions from object detection were compared to information extracted from the PDF version of the page or from specialized classifiers.
This seems to be applied in most cases to the title and text categories, which significantly improve the performance of previously reported results.

\begin{table}[htpb!]
\begin{center}
\begin{tabular}{|l|c|c|c|c|c|c|}
\hline
Team Name    &Text&Title&List&Table&Figure&Overall \\
\hline
Davar-Lab-OCR& \textbf{0.9838} & \textbf{0.9607} & 0.9680 & 0.9735 & 0.9804 & \textbf{0.9733} \\	
TAL          & 0.9823 & 0.9420 & \textbf{0.9700} & \textbf{0.9775} & \textbf{0.9833} & 0.9710 \\	
Simo         & 0.9810 & 0.9536 & 0.9636 & 0.9738 & 0.9796 & 0.9703 \\
BIT-VR Lab   & 0.9778 & 0.9270 & 0.9645 & 0.9762 & 0.9816 & 0.9654 \\	
IOD          & 0.9774 & 0.9251 & 0.9620 & 0.9773 & 0.9814 & 0.9647 \\	
\begin{CJK}{UTF8}{bsmi}小试牛刀\end{CJK}& 0.9797 & 0.9515 & 0.9575 & 0.9635 & 0.9709 & 0.9646 \\	
JHL          & 0.9774 & 0.9245 & 0.9620 & 0.9754 & 0.9814 & 0.9642 \\	
\begin{CJK}{UTF8}{bsmi}刷不动了\end{CJK}& 0.9778 & 0.9248 & 0.9634 & 0.9734 & 0.9803 & 0.9639 \\	
SRK           & 0.9767 & 0.9200 & 0.9599 & 0.9737 & 0.9800 & 0.9621 \\
\hline
\end{tabular}
\caption{Task A results}
\label{tab:task-a-results}
\end{center}
\end{table}

\subsection{Systems description}

These are the descriptions of the top systems provided by the participants for Task A\footnote{Not all descriptions for the top systems were provided.}.


\subsubsection{Team: Davar-Lab-OCR, Hikvision Research Institute}

The system is built based on a multi-modal Mask-RCNN-based object detection framework. For a document, we make full use of the advantages from vision and semantics, where the vision is introduced in the form of document image, while semantics (texts and positions) is directly parsed from PDF. We adopt a two-stream network to extract modality-specific visual and semantic features. The visual branch processes document image and semantic branch extracts features from text embedding maps (text regions are filled with the corresponding embedding vectors, which are learned from scratch). The features are fused adaptively as the complete representation of document, and then are fed into a standard object detection procedure.

To further improve accuracy, model ensemble technique is applied. Specifically, we train two large multimodal layout analysis models (a. ResNeXt-101-Cascade DCN Mask RCNN; b. ResNeSt-101-Cascade Mask RCNN), and inference the models in several different scales. The final results are generated by a weighted bounding-boxes fusion strategy. The code and related paper will be published in~\url{https://davar-lab.github.io/news.html}.

\subsubsection{Team: Tomorrow Advancing Life (TAL)}

TAL\footnote{http://www.100tal.com/about.html} used HTC (Hybrid Task Cascade for Instance Segmentation) as the baseline, which is an improved version of cascade mask rcnn. We first used some general optimization: 

(1) carefully designed the ratio of anchor;

(2) add deformable convolution module and global context block to the backbone;

(3) replace FPN with PAFPN;

(4) extract multi-level features instead of one-level features; (5) adopt IOU-balanced sampling to make the training samples more representative.

To tackle the difficulty of precise localization, we use two methods:

(1) we implement the algorithm SABL (Side-Aware Boundary Localization), where each side of the bounding box is respectively localized with a dedicated network branch;

(2) we train an expert model for the 'title' category to further improve the localization precision

In the post-processing stage, a classification model and self-developed text line detection model are used to solve the problem of missing detection in specific layout. In order to solve the problem of false detection of non target text, LayoutLM\footnote{\url{https://github.com/microsoft/unilm/tree/master/layoutlm}} is used to classify each line of text and remove the non target class.

At last, we ensemble multiple backbone models such as resnest200, resnext101, etc, and set different nms threshold for different categories.
References\footnote{LayoutLM: \url{https://github.com/microsoft/unilm/tree/master/layoutlm}}
\footnote{mmdetection:\url{https://github.com/open-mmlab}}.

\subsubsection{Team: Simo, Shanghai Jiao Tong University}

We treat the document layout analysis as an object detection task, and achieve it based on the framework of mmdetection. We first train a baseline model (Mask-RCNN). Afterwards, we improved our model from the following aspects:

1. \textit{Annotations}: We find that for the “text” category, some samples in the train dataset are unannotated, which leads to low recall of this category. Thus we design heuristic strategies to replenish the annotations in the training dataset, which can increase the overall AP on category of “text”.

2. \textit{Large models}: To improve performance, the network is trained based on a large backbone (ResNet-152), together with GCB and DCN blocks, which can improve our performance largely.

3. \textit{Results refinement}: For categories of “text” and “title”, we use the coordinates extracted from the PDF to refine the final results. Specifically, we parse the text line coordinates through PDFMiner, and refine the layout prediction (large box) using the above line coordinates.

4. \textit{Model ensemble}: Finally, we use model ensemble techniques to ensemble the above results as our final result.

\subsubsection{Team: SRK}

Our final solution is based on the ensemble of Mask Cascade R-CNN with ResNeSt-50 FPN backbone. First model was used for "Title" detection (small objects) and the second one for detection entities of other classes: "Text", "List", "Figure" and "Table". There were no any image augmentation techniques during models train and inference. Inference optimization was done by choosing NMS threshold parameter. The best result was obtained with a value of 0.9. We've used Detectron2 library for implementation and checking of our models. This solution is a continuation of our previous research on Document Layout Analysis problem, published in~\cite{grigoryev2021ivtov}.

\subsubsection{Team: BIT-VR Lab}
In this work, our base detection method follows the two-stage framework of DetectoRS that employs HTC branch to make full use of instance and segmentation annotation to enhance the feature flow in the feature extraction stage. We train a series of CNN models based on this method with different backbones, larger input image scales, customized anchor size, various loss functions,  rich data augmentation and soft-NMS method. More specifically, we use NAS technique to obtain optimal network architecture and optimal parameter configuration. Another technique is that we use OHEM to make training more effective and efficient and improve the detection accuracy of difficult samples like the “Title” category.

Besides, we trained Yolo-v5x model as our one-stage objection detection method, and CenteNet2 to take advantage of different characteristics in both one-stage and two-stage methods. To obtain the final ensemble detection results, we combine three different network frameworks as above and different multi-scale testing approaches with specific ensemble strategy. 
 
\section{Task B - Table Recognition}

Information in tabular format is prevalent in all sorts of documents. Compared to natural language, tables provide a way to summarize large quantities of data in a more compact and structured format. Tables provide as well a format to assist readers with finding and comparing information. 
This competition aims to advance the research in automated recognition of tables in unstructured formats. 

Participants of this task need to develop a model that can convert images of tabular data into the corresponding HTML code, which follows the HTML table representation from PubMed Central. The HTML code generated by the task participants should correctly represent the structure of the table and the content of each cell. HTML tags that define the text style including bold, italic, strike through, superscript, and subscript should be included in the cell content.
The HTML code does NOT need to reconstruct the appearance of tables such as border lines, background color or font, font size, or font color.
The competition site is available from~\footnote{\url{Task B website: https://aieval.draco.res.ibm.com/challenge/40/overview}}.

\subsection{Related work}

There are other table recognition challenges, which are mainly organized at the International Conference on Document Analysis and Recognition (ICDAR). ICDAR 2013 Table Competition is the first competition on table detection and recognition~\cite{gobel2013icdar}. A total of 156 tables are included in ICDAR 2013 Table Competition for evaluation of table detection and table recognition methods; however, no training data is provided. ICDAR 2019 Competition on Table Detection and Recognition provides training, validation, and test samples (3,600 in total) for table detection and recognition~\cite{gao2019competition}. Two types of documents, historical handwritten and model programmatic, are offered in image format. The ICDAR 2019 competition includes three tasks: 1) identifying table regions; 2) recognizing table structure with given table regions; 3) recognizing table structure without given table regions. The ground truth only includes the bounding box of table cell, without the cell content.

Our Task B competition proposed a more challenging task: the model needs to recognize both the table structure and the cell content of a table solely relying on the table image. In another word, the model needs to infer the tree-structure of the table and the properties (content, row-span, column-span) of each leaf node (table header/body cells). In addition, we do not provide intermediate annotations of cell position, adjacency relations, or row/column segmentation, which are needed to train most of the existing table recognition models. We only provide the final results of the tree representation for supervision. We believe this will motivate participants to develop novel models for image-to-structure mapping.

\subsection{Data}

This task used the PubTabNet dataset (v2.0.0)\footnote{\url{https://github.com/ibm-aur-nlp/PubTabNet}}~\cite{zhong2019image}. PubTabNet contains over 500k training samples and 9k validation samples, of which the ground truth HTML code, and the position of non-empty table cells are provided. Participants can use the training data to train their model and the validation data for model selection and hyper-parameter tuning. The 9k+ final evaluation set (image only, no annotation) was released 3 days before the competition ended for the Final Evaluation Phase. Participants submitted their results on this set in the final phase.

Submissions were evaluated using the TEDS (Tree-Edit-Distance-based Similarity) metric\footnote{\url{https://github.com/ibm-aur-nlp/PubTabNet/tree/master/src}}~\cite{zhong2019image}.
TEDS measures the similarity between two tables using the tree-edit distance proposed in~\cite{pawlik2016tree}.
The cost of insertion and deletion operations is 1.
When the edit is substituting a node $n_o$ with $n_s$, the cost is 1 if either $n_o$ or $n_s$ is not td. When both $no$ and $n_s$ are td, the substitution cost is 1 if the column span or the row span of $n_o$ and $n_s$ is different. Otherwise, the substitution cost is the normalized Levenshtein similarity~\cite{levenshtein1966binary} (in $[0, 1]$) between the content of $n_o$ and $n_s$. Finally, TEDS between two trees is computed as

\begin{equation}
TEDS(Ta, Tb) = 1 - \frac{EditDist(Ta, Tb)}{max(|Ta|, |Tb|)}
\end{equation}

where EditDist denotes tree-edit distance, and $|T|$ is the number of nodes in T. The table recognition performance of a method on a set of test samples is defined as the mean of the TEDS score between the recognition result and ground truth of each sample.

The competition had three phases. The Format Verification Phase spanned the whole competition, for participants to verify if their results file met our submission requirements with the mini development set that we provided. The Development Phase spanned from the beginning of the competition to 3 days before the competition ended. In this phase, participants could submit results on the test samples to verify their model. The Final Evaluation Phase run in the final 3 days of this competition. Participants could submit the inference results on the final evaluation set in this phase. Final ranking and winning teams were decided by the performance in the Final Evaluation Phase.
Table~\ref{tab:task-b-data} shows the size of the different data sets used in the different Task B phases.

\begin{table}[htpb!]
\begin{center}
\begin{tabular}{|l|r|l|}
\hline
Split           &Size   &Phase \\
\hline
Training        &500,777&N/A \\
Development     &9,115  &N/A \\
Mini development&20     &Format Verification Phase \\
Test            &9,138  &Development \\
Final evaluation&9,064  &Final evaluation \\
\hline
\end{tabular}
\end{center}
\label{tab:task-b-data}
\caption{Task B data set statistics}
\end{table}

\subsection{Results}

For Task B, we had 30 submissions from 30 teams for the Final Evaluation Phase.
Top 10 systems ranked using their TEDS performance on the final evaluation set are shown in table~\ref{tab:task-b-results}.
Due to a problem with the final evaluation data set, bold tags $<$b$>$ where not considered in the evaluation.

The first four systems have similar performance, while we see a more significant different thereafter.
As it is shown in the description of the systems, they rely on the combination of several components that identify relevant components from table images and then compose them.
The performance is better than compared to previously reported result of 91 in the TEDS metric using an image to sequence approach~\cite{zhong2019publaynet}.
In~\cite{zhong2019publaynet}, the data set is comparable to the test set of this competition and was derived as well from PubMed Central. 

\begin{table}[htpb!]
\begin{center}
\begin{tabular}{|l|c|c|c|}
\hline
Team Name&TEDS Simple & TEDS Complex & TEDS all \\
\hline
Davar-Lab-OCR	   & 97.88 & 94.78 & \textbf{96.36} \\
VCGroup	           & \textbf{97.90} & 94.68 & 96.32 \\
XM	               & 97.60 & \textbf{94.89} & 96.27 \\
YG	               & 97.38 & 94.79 & 96.11 \\
DBJ	               & 97.39 & 93.87 & 95.66 \\
TAL                & 97.30 & 93.93 & 95.65 \\
PaodingAI	       & 97.35 & 93.79 & 95.61 \\
anyone	           & 96.95 & 93.43 & 95.23 \\
LTIAYN	           & 97.18 & 92.40 & 94.84 \\
\hline
\end{tabular}
\caption{Task B top TEDS results. The overall result (TEDS all) is decompose into simple and complex tables~\cite{zhong2019image}}
\label{tab:task-b-results}
\end{center}
\end{table}

\subsection{Systems description}

These are the descriptions of the top systems provided by the participants for Task B\footnote{Not all descriptions for the top systems were provided.}.


\subsubsection{Team: Davar-Lab-OCR, Hikvision Research Institute}

The table recognition framework contains two main processes: table cells generation and structure inference\footnote{Davar-Lab-OCR paper and source code: \url{https://davar-lab.github.io}.}.

(1) Table cells generation is built based on the Mask-RCNN detection model. Specifically, the model is trained to learn the row/column aligned cell-level bounding boxes with corresponding mask of text content region. We introduce the pyramid mask supervision and adopt a large backbone of HRNet-W48 Cascade Mask RCNN to obtain the reliable aligned bounding boxes. In addition, we train a single-line text detection model with an attention-based text recognition model to provide the OCR information. This is simply achieved by selecting the instances that only contain single-line text. We also adopt multi-scale ensemble strategy on both the cell and single-line text detection models to further improve performance.

(2) In the structure inference stage, the bounding boxes for cells can be horizontally/vertically connected according to their alignment overlaps. The row/column information is then generated via a Maximum Clique Search process, during which empty cells can be easily located.   

To handle some special cases, we train another table detection model to filter out text not belonging to the table.

\subsubsection{Team: VCGroup}

In our method~\cite{he2021ICDAR,lu2019master,ye2021ICDAR}\footnote{VCGroup Github repo: \url{https://github.com/wenwenyu/MASTER-pytorch}}, we divide the table content recognition task into four sub-tasks: table structure recognition, text line detection, text line recognition and box assignment. Our table structure recognition algorithm is customized based on MASTER, a robust image text recognition algorithm. PSENet is used to detect each text line in the table image. For text line recognition, our model is also built on MASTER. Finally, in the box assignment phase, we associated the text boxes detected by PSENet with the structure item reconstructed by table structure prediction, and fill the recognized content of the text line into the corresponding item. Our proposed method achieves a 96.84\% TEDS score on 9,115 validation samples in the development phase, and a 96.32\% TEDS score on 9,064 samples in the Final Evaluation Phase. 

\subsubsection{Team: Tomorrow Advancing Life(TAL)}

The TAL system consists of two schemes:

1. Rebuild table structure through 5 detection models, which are table head-body detection, row detection, column detection, cell detection and text-row detection. Mask R-CNN is selected as the baseline for these 5 detection models, with targeted optimization for different detection tasks. In the recognition part, the results of cell detection and text-row detection are inputted into the CRNN model to get the recognition result corresponding to each cell.

2. The restoration of table structure is treated as an img2seq problem. To shorten the decoding length, we replace every cell content with different numbers. The numbers are obtained from  text-row detection results. Then we use CNN to encode the image and use a transformer model to decode the structure of the table. The corresponding text-line content can then be obtained by using the CRNN model.

The above two schemes can be used to get the complete table structure and content recognition results. We have a set of selection rules, which combine the advantages of both schemes, to output the one best final result.

\subsubsection{Team: PaodingAI, Beijing Paoding Technology Co., Ltd}

PaodingAI's system is divided into three main parts: text block detection, text block recognition and table structure recognition. The text block detector is trained by the Detectors\_cascade\_rcnn\_r50\_2x model provided by MMDetection. The text block recognizer is trained by the SAR\_TF \footnote{\url{https://github.com/Pay20Y/SAR\_TF}} model. Table structure recognizer is our own implementation of the model proposed in~\cite{tensmeyer2019deep}. In addition to the above model, we also use rules and a simple classification model to process $<$thead$>$, $<$b$>$, and blank characters. Our system is not an end-to-end model and does not use an integrated approach.

\subsubsection{Team: Kaen Context, Kakao Enterprise}\footnote{Company located in Seongnam-si, Gyeonggi-do, South Korea}

To resolve the problem of table recognition in an efficient way, we use the 12-layer decoder-only linear transformer architecture~\cite{katharopoulos2020transformers}.

Data preparation: We use RGB images without rescaling as input conditions and the merged HTML code is used as target text sequences. We reshape a table image into a sequence of flattened patches with shape (N, 8*8*3), where 8 is the width and height of each image patch, and N is the number of patches. Then, we map the image sequence to 512 dimensions with a linear projection layer. The target text sequence is converted into a 512-dimensional embedding through an embedding layer and appended at the end of the projected image sequence. Finally, we add different positional encodings to the text and image sequences to allow our model to distinguish them.

Training: The concatenated image-text sequence is used as the input of our model and the model is trained by the cross-entropy loss under the teacher forcing algorithm.

Inference: The outputs of our model are sampled with beam search (beam=32).

\section{Conclusions}

We have proposed two tasks for document understanding using large data sets derived from PubMed Central for the training and evaluation of participant systems.
These tasks address two important problems, understanding document layouts and table identification, including both table border and cell structure.

We had a large participation for both tasks, which was quite significant for Task A with 281 submissions from 78 teams.
Results from top participant submissions significantly improve the performance of previously reported results.

Results from both tasks show an impressive performance and opens the possibility for high performance practical applications.
There are still some aspects to improve from Task A, such as a better identification of titles, and better processing of complex tables in Task B.
Both tasks have used a data set derived from scientific literature. The generated data sets are quite large and diverse in the formats in which the information is represented. This diversity should help using the trained models in other domains, which could be evaluated using new data sets generated for other domains such as FinTabNet~\cite{zheng2021global} for the financial domain.

\section{Acknowledgements}

We would like to thank Sundar Saranathan for his help with the competition system.

We would like to thank the US NIH/National Library of Medicine for making available the data sets used in this competition.

%
%
%
\bibliographystyle{splncs04}
\bibliography{bibliography}

\begin{thebibliography}{10}
\providecommand{\url}[1]{\texttt{#1}}
\providecommand{\urlprefix}{URL }
\providecommand{\doi}[1]{https://doi.org/#1}

\bibitem{antonacopoulos2009realistic}
Antonacopoulos, A., Bridson, D., Papadopoulos, C., Pletschacher, S.: A
  realistic dataset for performance evaluation of document layout analysis. In:
  2009 10th International Conference on Document Analysis and Recognition. pp.
  296--300. IEEE (2009)

\bibitem{clausner2017icdar2017}
Clausner, C., Antonacopoulos, A., Pletschacher, S.: Icdar2017 competition on
  recognition of documents with complex layouts-rdcl2017. In: 2017 14th IAPR
  International Conference on Document Analysis and Recognition (ICDAR).
  vol.~1, pp. 1404--1410. IEEE (2017)

\bibitem{clausner2015enp}
Clausner, C., Papadopoulos, C., Pletschacher, S., Antonacopoulos, A.: The enp
  image and ground truth dataset of historical newspapers. In: 2015 13th
  International Conference on Document Analysis and Recognition (ICDAR). pp.
  931--935. IEEE (2015)

\bibitem{gao2019competition}
Gao, L., Huang, Y., Li, Y., Yan, Q., Fang, Y., Dejean, H., Kleber, F., Lang,
  E.M.: {ICDAR} 2019 competition on table detection and recognition. In: 2019
  International Conference on Document Analysis and Recognition (ICDAR). pp.
  1510--1515. IEEE (Sep 2019). \doi{10.1109/ICDAR.2019.00166}

\bibitem{gobel2013icdar}
G{\"o}bel, M., Hassan, T., Oro, E., Orsi, G.: {ICDAR} 2013 table competition.
  In: 2013 12th International Conference on Document Analysis and Recognition.
  pp. 1449--1453. IEEE (2013)

\bibitem{grigoryev2021ivtov}
Grygoriev, A., Degtyarenko, I., Deriuga, I., Polotskyi, S., Melnyk, V.,
  Zakharchuk, D., Radyvonenko, O.: {HCRNN}: A novel architecture for fast
  online handwritten stroke classification. In: Proc. of Int. Conf. on Document
  Analysis and Recognition (2021)

\bibitem{he2021ICDAR}
He, Y., Qi, X., Ye, J., Gao, P., Chen, Y., Li, B., Tang, X., Xiao, R.:
  Pingan-vcgroup’s solution for icdar 2021 competition on scientific table
  image recognition to latex. arXiv  (2021)

\bibitem{katharopoulos2020transformers}
Katharopoulos, A., Vyas, A., Pappas, N., Fleuret, F.: Transformers are rnns:
  Fast autoregressive transformers with linear attention. In: International
  Conference on Machine Learning. pp. 5156--5165. PMLR (2020)

\bibitem{levenshtein1966binary}
Levenshtein, V.I.: Binary codes capable of correcting deletions, insertions,
  and reversals. In: Soviet physics doklady. vol.~10, pp. 707--710. Soviet
  Union (1966)

\bibitem{lu2019master}
Lu, N., Yu, W., Qi, X., Chen, Y., Gong, P., Xiao, R., Bai, X.: Master:
  Multi-aspect non-local network for scene text recognition. Pattern
  Recognition  (2021)

\bibitem{pawlik2016tree}
Pawlik, M., Augsten, N.: Tree edit distance: Robust and memory-efficient.
  Information Systems  \textbf{56},  157--173 (2016)

\bibitem{staar2018corpus}
Staar, P.W., Dolfi, M., Auer, C., Bekas, C.: Corpus conversion service: A
  machine learning platform to ingest documents at scale. In: Proceedings of
  the 24th ACM SIGKDD International Conference on Knowledge Discovery \& Data
  Mining. pp. 774--782 (2018)

\bibitem{tensmeyer2019deep}
Tensmeyer, C., Morariu, V.I., Price, B., Cohen, S., Martinez, T.: Deep
  splitting and merging for table structure decomposition. In: 2019
  International Conference on Document Analysis and Recognition (ICDAR). pp.
  114--121. IEEE (2019)

\bibitem{ye2021ICDAR}
Ye, J., Qi, X., He, Y., Chen, Y., Gu, D., Gao, P., Xiao, R.: Pingan-vcgroup's
  solution for icdar 2021 competition on scientific literature parsing task b:
  Table recognition to html. arXiv  (2021)

\bibitem{zheng2021global}
Zheng, X., Burdick, D., Popa, L., Zhong, X., Wang, N.X.R.: Global table
  extractor (gte): A framework for joint table identification and cell
  structure recognition using visual context. In: Proceedings of the IEEE/CVF
  Winter Conference on Applications of Computer Vision. pp. 697--706 (2021)

\bibitem{zhong2019image}
Zhong, X., ShafieiBavani, E., Yepes, A.J.: Image-based table recognition: data,
  model, and evaluation. arXiv preprint arXiv:1911.10683  (2019)

\bibitem{zhong2019publaynet}
Zhong, X., Tang, J., Yepes, A.J.: Publaynet: largest dataset ever for document
  layout analysis. In: 2019 International Conference on Document Analysis and
  Recognition (ICDAR). pp. 1015--1022. IEEE (2019)

\end{thebibliography}

\end{document}